\documentclass[11pt]{iopart}
\usepackage{bm}
\usepackage{hyperref}
\usepackage{amssymb}
\usepackage{amsopn}
\usepackage{slashed}
\RequirePackage{ifpdf}

\begin{document}

	\title{Laboratory Tests of the Galileon}
	
	\author{Philippe Brax$^{1}$,  Clare Burrage$^2$ and Anne-Christine Davis$^3$ }
	\vspace{3mm}
	
	\address{$^1$ Institut de Physique Th\'{e}orique,
	  CEA, IPhT, CNRS, URA2306, F-91191 Gif-sur-Yvette c\'{e}dex,
	  France \\[3mm]
	$^2$ 	  D\'epartement de Physique Th\'eorique and Center for Astroparticle Physics,
Universit\'e de Gen\`eve, 24 quai Ansermet, CH--1211 Gen\`eve 4,
Switzerland \\[3mm]
	$^3$ Department of Applied Mathematics and Theoretical Physics,\\
Centre for Mathematical Sciences, Cambridge CB3 0WA, United Kingdom
	 	}

	\vspace{3mm}

	\eads{\mailto{philippe.brax@cea.fr},
	\mailto{clare.burrage@unige.ch},
	\mailto{A.C.Davis@damtp.cam.ac.uk}}
	
	\begin{abstract}
The Galileon model is a ghost free scalar effective field theory containing higher derivative terms that are protected by the Galileon symmetry.  The presence of a Vainshtein screening mechanism allows the scalar field to couple to matter without mediating unacceptably large fifth forces in the solar system.  We describe how laboratory measurements of the Casimir effect and possible deviations from Newtonian gravity can be used to search for Galileon scalar fields.  Current experimental measurements are used to bound a previously unconstrained combination of Galileon parameters.

	\end{abstract}
	\maketitle

	\section{Introduction}
	\label{sec:introduction}

The theory of General Relativity has successfully passed a wide range of tests of its validity over a huge variety of distance scales; from millimetre scales in the laboratory, through solar system scales to the Megaparsec scales of galaxies and galaxy clusters.  However on the very largest scales in the universe something appears to go wrong.  If the evolution of the universe is governed by general relativity, and if there is no exotic matter in the universe, then the expansion of the universe should be decelerating.  However, despite all expectations, a variety of different large scale observations have been combined to show that the expansion of the universe is accelerating.

The simplest explanation of this observation is a straightforward modification of the Einstein equations to include a cosmological constant term.  However reconciling theoretical predictions with observations of its magnitude requires a massive fine tuning of up to 120 orders of magnitude.  If  this fine tuning issue makes a cosmological constant explanation unacceptable we must either include a matter field with fundamentally new properties or make a more dramatic modification of the theory of gravity.

In practice, and excepting extreme environments such as black holes, the two possibilities of modifying gravity or introducing new matter often reduce to the same thing in four dimensions:  The introduction of new, scalar degrees of freedom either in the matter or in the gravitational sector of the theory. The introduction of new scalar fields is not without its own issues; if they couple to matter it is necessary to explain why no signs of a scalar force  have been seen in laboratory and solar system measurements.

For the scalar force to have significant effects on large scales, but negligible effects on small scales it seems highly probable that its behaviour must be non-linear.  One such mechanism for achieving this, named after Vainshtein,  allows the coupling between the scalar field and matter to become weak in the presence of matter sources \cite{Vainshtein:1972sx}
\footnote{Other screening mechanisms are available, for example the chameleon \cite{Khoury:2003rn} and symmetron models \cite{Hinterbichler:2010es}.}.
The Vainshtein mechanism was originally proposed to allow the helicity-zero mode of a massive four dimensional  graviton to decouple in the presence of sources, thus resolving the vDVZ discontinuity \cite{Vainshtein:1972sx,Deffayet:2001uk}.  Of particular interest for this paper is that the Vainshtein mechanism is exhibited by a new class of scalar field theories dubbed `Galileon' \cite{Nicolis:2008in}.  Galileon models  can be related to theories of massive gravity \cite{deRham:2010ik}, or those with extra dimensions \cite{Dvali:2000hr,deRham:2010eu,VanAcoleyen:2011mj}, but for the purpose of this paper it suffices only to consider them as four dimensional effective field theories.

A Galileon scalar field theory is a non-linear model with two characteristic properties:  (i) It obeys a generalised form of the Galileon symmetry $\pi\rightarrow \pi + c + b^{\mu}x_{\mu}$. (ii) The equations of motion arising from the Galileon Lagrangian are second order in derivatives, meaning that the theory may   contain higher derivative  terms in the Lagrangian but remains free of ghost-like instabilities.  In four dimensions, there are only five possible operators with these properties \cite{Nicolis:2008in}.

The phenomenology \cite{Burrage:2010rs,Babichev:2010kj,Wyman:2011mp} and cosmology \cite{Silva:2009km,Gannouji:2010au,Nesseris:2010pc,DeFelice:2010as} of the Galileon have been extensively studied.  It has also been considered as a possible candidate for inflation where it gives rise to distinctive observational signatures \cite{Burrage:2010cu,Creminelli:2010qf,Mizuno:2010ag}, related inflationary models known as G-inflation that do not respect the Galileon symmetry were proposed in \cite{Kobayashi:2010cm}.  The Galileon has also been promoted  to a covariant theory \cite{Deffayet:2009wt,Deffayet:2009mn,Chow:2009fm}, and  to theories of multiple fields \cite{Deffayet:2010zh,Padilla:2010de,Padilla:2010tj,Hinterbichler:2010xn}. The generalisation of the Galileon symmetry to curved space backgrounds has also been studied \cite{Goon:2011qf,Burrage:2011bt}.

In four dimensions the Galileon Lagrangian contains four independent terms each entering with its own coefficient\footnote{In \cite{Nicolis:2008in} a tadpole is included as a fifth Galileon term.  We neglect this here as simply a rescaling of the cosmological constant.}.  Therefore at least four observations are required to determine or constrain the whole of the Galileon parameter space.  In Section \ref{sec:llr} we review the constraints we have so far from Lunar Laser Ranging, which restricts the perturbation due to the Galileon on the precession of the perihelion of the moon.  In what follows we discuss how Laboratory experiments can be used to constrain an independent combination of Galileon parameters, we focus in particular on experiments which study forces between parallel plates.  In Section \ref{sec:plates} we compute the form of the Galileon field around parallel plates in a laboratory.  Then in Section \ref{sec:exp} we apply the constraints from  the E\"{o}t-Wash experiments \cite{Kapner:2006si} to search for deviations from Newtonian gravity, we also discuss whether experimental tests of the Casimir force are suitable to constrain the Galileon.  We begin in the next Section with a review of the Galileon model.

\section{A Review of the Galileon}
\label{sec:review}

In this paper we restrict our study to the simplest realisation of the Galileon model that contains only a single scalar field in flat space.  This is described by the following Lagrangian.
\begin{equation}
\mathcal{L}=\frac{1}{2}c_2(\partial \pi)^2 + \frac{1}{2}c_3 \Box \pi (\partial \pi)^2 +c_4\mathcal{L}_4(\pi)+c_5\mathcal{L}_5(\pi)+\pi T\;,
\label{eq:lag}
\end{equation}
where   $\Box\pi=\partial_{\mu}\partial^{\mu}\pi$, $T$ is the trace of the energy momentum tensor and   the terms $\mathcal{L}_4(\pi)$ and $\mathcal{L}_5(\pi)$ are given by
   \begin{eqnarray}
   \mathcal{L}_4(\pi)&=& -\frac{1}{4}\left[(\Box\pi)^2\partial\pi\cdot\partial\pi-2 (\Box\pi)^2\partial\pi\cdot\Pi\cdot\partial\pi\right. \nonumber \\
   & & \;\;\;\; \left.-(\Pi\cdot\Pi)(\partial\pi\cdot\partial\pi)+2\partial\pi\cdot \Pi\cdot \Pi \cdot \partial\pi\right]\;,\\
   \mathcal{L}_5(\pi)&=&-\frac{1}{5}\left[(\Box\pi)^3\partial\pi\cdot\partial\pi-3(\Box\pi)^2\partial\pi\cdot\Pi\cdot\partial\pi-3\Box\pi(\Pi\cdot\Pi)(\partial\pi\cdot\partial\pi)\right.\nonumber\\
   & & \;\;\;\; +6(\Box\pi)\partial\pi\cdot\Pi\cdot\Pi\cdot \partial\pi+2(\Pi\cdot\Pi\cdot\Pi)(\partial\pi\cdot\partial\pi)\nonumber\\
   & & \;\;\;\; \left.+3(\Pi\cdot\Pi)\partial\pi\cdot \Pi\cdot\partial\pi-6\partial\pi\cdot\Pi\cdot\Pi\cdot\Pi\cdot\partial\pi\right]\;,
      \end{eqnarray}
where $\Pi^{\nu}_{\mu}=\partial_{\mu}\partial_{\nu}\pi$, and contraction of indices is implied.
    The $c_i$ are arbitrary  coefficients, with dimensions of mass to the power $2(3-i)$.  The energy scales in these coefficients do not all necessarily have to be the same.  Motivation for allowed mass hierarchies comes from the very different origins of the coefficients in the probe brane theories that give rise to the Galileon \cite{deRham:2010eu}, and also arise in multi-Galileon theories with softly broken internal symmetries \cite{Hinterbichler:2010xn}.  These energy scales are also not necessarily directly connected to the cut off energy scale of the low energy effective theory.  They may be intermediate mass scales in the theory, indeed this is the case in the DGP and massive gravity completions of the Galileon model \cite{deRham:2010ik,Dvali:2000hr}.  Therefore we can consider energy scales for the coefficients as low as $\sim 1/(1000 \mbox{ km})$ whilst still trusting the form of the Galileon Lagrangian on the millimetre scales of laboratory measurements, as long as we bear in mind that the assumed cut off of the theory must be an energy scale higher than $\sim 1/\mbox{ mm}$ \footnote{We would like to thank Andrew Tolley for bringing this  to our attention.}.
    Equation (\ref{eq:lag}) is the form of the Galileon action proposed in \cite{Nicolis:2008in}, the purely scalar part of the Lagrangian respects the Galileon symmetry, $\pi\rightarrow \pi +c+b^{\mu} x_{\mu}$ and is  defined up to total derivative terms which are irrelevant in flat space.

    The Galileon model was originally studied as an explanation of the accelerated expansion of the universe, where the background scalar field has a particular `self-accelerated' form \cite{Nicolis:2008in}.  In this case the scalar field profiles sourced by massive objects in the universe are small perturbations of this background.  The form of the Lagrangian for these perturbations is exactly that of Equation (\ref{eq:lag}), and the effect of the background is a constant rescaling of the coefficients $c_i$ in a way which is described in detail in \cite{Nicolis:2008in}.

The equation of motion for the scalar field obtained from the Lagrangian (\ref{eq:lag}) is
\begin{eqnarray}
c_2\Box\pi+c_3\left[(\Box\pi)^2-(\partial_{\mu}\partial_{\nu}\pi)^2\right]&& \label{eq:eom}\\
+c_4[(\Box\pi)^3-3\Box\pi(\partial_{\mu}\partial_{\nu}\pi)^2+2(\partial_{\mu}\partial_{\nu}\pi)^3]& &\nonumber\\
+c_5[(\Box\pi)^4-6(\Box\pi)^2(\partial_{\mu}\partial_{\nu}\pi)^2+8(\Box\pi)(\partial_{\mu}\partial_{\nu}\pi)^3& &\nonumber \\
\;\;\;\;+3[(\partial_{\mu}\partial_{\nu}\pi)^2]^2-6(\partial_{\mu}\partial_{\nu}\pi)^4]& =&-T\nonumber\;.
\end{eqnarray}
We will shortly recall some properties  of this equation when solved in the solar system and then consider the situation of laboratory experiments. 

\subsection{Possible UV Origins of the Galileon}
The Galileon model can be obtained as the effective four dimensional description of scalar degrees of freedom arising in certain extra dimensional theories, and in massive gravity.  We briefly summarize these connections, and the values of the parameters $c_i$ obtained.
\begin{itemize}
\item {\bf DGP.} The prototype  Galileon theory is the DGP brane world scenario \cite{Dvali:2000hr}: A four dimensional brane moves in a five dimensional bulk space, where there is both  a five-dimensional Ricci scalar in the bulk geometry, and  a four-dimensional Ricci scalar on the brane.  There are two energy scales in this system, one becomes the four dimensional Planck mass, and  the other, to agree with observations, must be fixed to be the measured value of the cosmological constant today $\Lambda \sim 10^{-3}\mbox{ eV}$.  This fixes $c_2 = 24 m_P^2\sim 10^{39}\mbox{ GeV}^2$,  $c_3 = 16 (m_P/\Lambda)^3\sim 10^{94}$ and $c_4=c_5=0$.

\item {\bf Probe brane worlds.}  If a four-dimensional probe brane is moving in a five dimensional bulk there is a scalar degree of freedom describing its position in the fifth dimension.  In the low energy four dimensional effective theory the action for this scalar degree of freedom has Galileon form \cite{deRham:2010eu}.  Then $c_2$ is related to the brane tension, and the higher coefficients $c_3$, $c_4$ and
$c_5$ receive contributions from the normalization of higher curvature invariants in the
five-dimensional brane action.

\item {\bf Massive Gravity.} A massless graviton in four dimensions has two degrees of freedom, but a massive one has five.  These can be decomposed into two tensor modes, two vector modes, and a scalar mode.  It is an unresolved question whether the graviton of the Standard Model has a mass, although if it does it must be smaller than the value of the Hubble scale today.  Care is needed to ensure that a theory of massive gravity does not contain ghost degrees of freedom, but in scenarios where these pathologies are absent the action for the scalar degree of freedom reduces to the Galileon form \cite{deRham:2010ik} with the following values of the parameters
\begin{eqnarray}
c_i&=&\tilde{c}_i\frac{m_P^i}{\Lambda_3^{3(i-2)}} \;\;\;\;\; 2\leq i\leq 4\;, \\
c_5&=&0\;.
\end{eqnarray}
Here $\Lambda_3=(m^2m_P)^{1/3}$ where $m$ is the mass of the graviton, and the $\tilde{c}_i$ are dimensionless order one coefficients.
\end{itemize}

\subsection{The Vainshtein Effect}
\label{sec:vainshtein}

The  Vainshtein effect, which screens the force mediated by the scalar field, occurs around static, spherically symmetric sources where the Galileon equation of motion (\ref{eq:eom}) becomes
\begin{equation}
\frac{1}{r^2}\frac{\partial}{\partial r}r^3\left[c_2\left(\frac{\pi^{\prime}}{r}\right)+2c_3\left(\frac{\pi^{\prime}}{r}\right)^2+2c_4\left(\frac{\pi^{\prime}}{r}\right)^3 \right]=M\delta^{(3)}(r)\;,
\label{eq:eomsphere}
\end{equation}
and we treat the  source of mass $M$ as a point like object situated at the origin, an approximation that is valid in the exterior of  the object.   The $\mathcal{L}_5$ operator does not contribute to the equation of motion as it is zero when evaluated on a static spherically symmetric profile.

Setting $c_3=c_4=0$ we recover the canonical result for the scalar force sourced by a point like object, $\pi^{\prime}=M/4\pi c_2 r^2$.
When the non-linearities are present, we parameterise deviations from the canonical result in terms of a dimensionless function $g(r)$, writing $\pi^{\prime}=(M/4\pi c_2 r^2)g(r)$.  The equation of motion is then an algebraic equation for $g$
\begin{equation}
g+\left(\frac{R_{\star}}{r}\right)^3g^2+\left(\frac{R_{2}}{r}\right)^6g^3=1\;,
\label{eq:eomg}
\end{equation}
where
\begin{eqnarray}
R_{\star}^3&=&c_3 M/2\pi c_2^2 \label{eq:rstar}\;,\\
R_2^6&=&M^2c_4/8\pi^2c_2^3\;,
\end{eqnarray}
 and there is an arbitrary constant of integration which we have set to zero to ensure that the force, which is proportional to $\pi^{\prime}$, vanishes at infinity.

For stable solutions $0\leq R_2<R_{\star}$ \cite{Nicolis:2008in}.
   If $R_{\star}\neq0$ then to avoid discontinuities in (\ref{eq:eomg}) we must impose  $g\rightarrow 0$ as $r\rightarrow 0$, such that  for $r<R_{\star}$ the function $g(r)$ falls below unity, and the scalar force is suppressed.  The radius within which the force is suppressed, $R_{\star}$ is known as the Vainshtein radius.

Within the Vainshtein radius the field has two different behaviours
\begin{itemize}
\item For $(R_2/R_{\star})^3R_2\ll r \ll R_{\star}$
\begin{equation}
\pi^{\prime}=\frac{M}{4\pi c_2 r^2}\left(\frac{r}{R_{\star}}\right)^{3/2}=\left(\frac{M}{8\pi c_3}\right)^{1/2}r^{-1/2}\;.
\label{eq:piprimeA}
\end{equation}

\item For $0\leq r \ll (R_2/R_{\star})^3R_2$
\begin{equation}
\pi^{\prime}=\frac{M}{4\pi c_2 r^2}\left(\frac{r}{R_{2}}\right)^{2}=\left(\frac{M}{8\pi c_4}\right)^{1/3}\;.
\label{eq:piprimeB}
\end{equation}
\end{itemize}

\subsection{Gravitationally Bound Systems}
Assuming that the unscreened force is at least as strong as gravity $c_2\leq m_P^2$ then a basic constraint comes from ensuring that systems, such as galaxies and clusters, which are gravitationally bound cosmologically do not feel order one deviations from general relativity.  It was shown in \cite{Burrage:2010rs} that this imposes
\begin{equation}
c_3\gtrsim 10^{118}\;.
\label{eq:gbs}
\end{equation}
This bound is obtained by requiring that the gravitationally bound systems must be well within the radius where the strength of the scalar force falls below that of the gravitational force.

\section{Lunar Laser Ranging}
\label{sec:llr}

More precise constraints on Galileon models come from Lunar Laser Ranging (LLR) experiments \cite{Dvali:2002vf}.  Even though the Galileon force is suppressed within the Vainshtein radius of the Earth it does not vanish and there is  a small modification to the Newtonian gravitational potential, $\Psi(r)=-GM/r$.  This  causes  a perturbation to the advance of the perihelion of the Moons orbit, something that is  very precisely measured by LLR \cite{2002nmgm.meet.1797W}.
We review here the constraints currently imposed by LLR, however we note that the calculation has currently only been performed to leading order, where the moon is treated as a test particle moving in the spherically symmetric field due to the Earth.  A full analysis will give small corrections to these constraints.

Generally if $\epsilon$ is the fractional change in the gravitational potential
\begin{equation}
\epsilon \equiv \frac{\delta\Psi}{\Psi}\;,
\end{equation}
the anomalous perihelion precession is \cite{Gruzinov:2001hp,Dvali:2002vf}
\begin{equation}
\delta\phi=\pi r \frac{\partial}{\partial r}\left[r^2\frac{\partial}{\partial r}\left(\frac{\epsilon}{r}\right)\right]\;.
\end{equation}

For the Galileon model we must ensure that the orbit of the moon lies within the Vainshtein radius of the Earth.  The maximum distance from the Earth to the Moon is $4.1\times 10^5 \mbox{ km}$, and the Vainshtein radius is larger than this if  $1.7\times 10^{22}\mbox{ GeV}^{-4}<c_3/c_2^2$.  The constraint on $c_3$  from gravitationally bound objects (\ref{eq:gbs}) ensures that this inequality is always satisfied. There are two regions of behaviour of the Galileon inside the Vainshtein radius:
\begin{itemize}
\item If the orbit of the moon lies in the range $(R_2/R_{\star})^3R_2\ll r \ll R_{\star}$, which requires $c_4^2/4c_3^3 <1.7\times 10^{22}\mbox{ GeV}^{-4}<c_3/c_2^2$, then
\begin{equation}
\epsilon=m_p^2 \left(\frac{r^3}{8\pi c_3 M}\right)^{1/2}\;,
\end{equation}
and the anomalous perihelion precession is
\begin{equation}
\delta\phi =\frac{3\pi m_P^2}{4}\left(\frac{r^3}{8\pi c_3 M}\right)^{1/2}\;.
\end{equation}

\item If the orbit of the moon lies in the range $0\leq r \ll (R_2/R_{\star})^3R_2$, which requires $c_4^2/c_3^3<6.8\times 10^{22}\mbox{ GeV}^{-4}$ then
\begin{equation}
\epsilon=\frac{m_p^2}{2} \left(\frac{1}{\pi c_4 M^2}\right)^{1/3}r^2\;,
\end{equation}
and the anomalous perihelion precession is
\begin{equation}
\delta\phi = m_P^2\left(\frac{\pi^2}{c_4 M^2}\right)^{1/2}\;.
\end{equation}
\end{itemize}

LLR measurements constrain $|\delta\phi| < 2.4 \times 10^{-11}$ implying that one of the following pairs of conditions must be satisfied
\begin{equation}
\begin{array}{cccc}
\mbox{either} & c_4^2/4c_3^3 <1.7\times 10^{22}\mbox{ GeV}^{-4}<c_3/c_2^2 & \mbox{and} &10^{120}<c_3\;,\\
\mbox{or} & c_4^2/c_3^3<6.8\times 10^{22}\mbox{ GeV}^{-4} & \mbox{and} &10^{190}\mbox{ GeV}^{-2}<c_4\;.
\end{array}
\end{equation}

\section{The Galileon in the Laboratory}
\label{sec:plates}
We consider two experiments that can be used to constrain the existence of the Galileon;  measurements of the Casimir effect and tests of gravity by the E\"{o}t-Wash collaboration.  Both experiments constrain the existence of new forces between two plates held close together.

Firstly we compute the  Galileon force around a single plate.  The plate has  density $\rho$ and thickness $\Delta$, and is  oriented perpendicular to the $z$-axis of our coordinate system  with its center  at $z=0$. If we approximate the plate as having infinite extent in the $x$ and $y$ directions the system becomes one-dimensional and a quick inspection of (\ref{eq:eom}) is sufficient to see that all of the non-linear terms will vanish.  There is no screening of the Galileon force around the plate, and the solution for the Galileon force is
\begin{equation}
\pi^{\prime}=\frac{\rho}{2c_2}\left\{\begin{array}{lc}
\Delta & r>\Delta/2 \\
2z & \Delta/2>r>-\Delta/2  \\
-\Delta & -\Delta/2>r
\end{array}\right.\;,
\label{eq:solplate}
\end{equation}
where we have imposed continuity of $\pi^{\prime}$ at the boundary of the plate, and $\pi^{\prime}(z)=-\pi^{\prime}(-z)$.

\subsection{At the surface of the Earth}
\label{sec:Earth}
Approximating the plate as a one-dimensional object  means that it is not surrounded by a region in which the non-linear terms in the equation of motion dominate and so the Galileon force due to the plates is not screened.  If there were nothing else in the universe a planar fifth force experiment would be extremely constraining for the Galileon.  However our universe is not quite that simple.  In particular we must take into account that the experiments are done on Earth so there is a  background  scalar field profile sourced by the Earth, and the plates of the experiment  source a small perturbation about this background.

The Vainshtein radius of the Earth can be computed from Eq. (\ref{eq:rstar}), and the constraints from gravitationally bound systems mean  that it is required to satisfy $R_{\star}\gtrsim 10^{18} \mbox{ cm}$.
The radius of the Earth is only $10^8 \mbox{ cm}$ and therefore the Earth lies well inside its Vainshtein radius.
Within the Vainshtein radius of the Earth we call  the background field profile  $\pi_{\oplus}(r)$, which can be determined from Equations (\ref{eq:piprimeA}) and (\ref{eq:piprimeB}).
The solution due to the plates of a laboratory experiment will be a small perturbation around this background.  We write $\pi(\vec{x})=\pi_{\oplus}(r)+\phi(z)$, where $\phi(z)$ is the perturbation due to the plates,  most conveniently expressed in terms of Cartesian coordinates, and the plates are aligned perpendicular to the $z$ axis.  We take the origin of both spherical and Cartesian coordinate systems to be at the center of the Earth, and relate the two systems in such a way that $r=z\cos\theta$.  By substituting $\pi(\vec{x})=\pi_{\oplus}(r)+\phi(z)$ into the equation of motion (\ref{eq:eom}), we can find the full equation of motion for $\phi$.  The only approximation we make is to assume that the extent of the experiment is always much less than the radius of the Earth so that everywhere inside the experiment $\theta\sim10^{-6}\ll1$.  Therefore whenever $\theta$ appears explicitly in the equation of motion for $\phi$ it is a good approximation to set $\cos\theta=1$ and $\sin\theta=0$.  After making this approximation the equation of motion becomes
\begin{equation}
\frac{d^2\phi}{dz^2}\left[c_2+4c_3\frac{\pi_{\oplus}^{\prime}}{r}+12c_4\left(\frac{\pi_{\oplus}^{\prime}}{r}\right)^2+32c_5\left(\frac{\pi^{\prime}_{\oplus}}{r^3}\right)^3\right]=-T\;.
\label{eq:eomlab}
\end{equation}
For convenience we define a function $Z(r)$ to be equal to the content of the square bracket in Eq. (\ref{eq:eomlab}). Over length scales much smaller than the radius of the Earth, including those of the experiments we consider, the variation in $Z$ is small $\Delta Z=|Z(R_{\oplus}+\epsilon)-Z(R_{\oplus})|\ll |Z(R_{\oplus})|$.
Therefore, over the extent of the experiment, $Z$ can be treated as constant with $r=R_{\oplus}$.
\begin{eqnarray}
Z_{\oplus}&=& c_2+4c_3\frac{\pi_{\oplus}^{\prime}}{R_{\oplus}}+12c_4\left(\frac{\pi_{\oplus}^{\prime}}{R_{\oplus}}\right)^2+32c_5\left(\frac{\pi^{\prime}_{\oplus}}{r_{\oplus}^3}\right)^3\;,\\
&=& \frac{R_{\oplus}}{\pi_{\oplus}^{\prime}}\left[\frac{2\rho_{\oplus}}{3}+8c_4\left(\frac{\pi^{\prime}_{\oplus}}{R_{\oplus}^3}\right)^3+32c_5\left(\frac{\pi^{\prime}_{\oplus}}{R_{\oplus}^3}\right)^4\right]\;,
\end{eqnarray}
where  the last line is obtained using the equation determining  the background field profile (\ref{eq:eomsphere}) to eliminate $c_3$.

When placed in the background field due to the Earth  the Galileon force due to a plate will have the same form as (\ref{eq:solplate}),  but with the coupling to matter rescaled, $c_2\rightarrow Z_{\oplus}$.  The exact size of $Z_{\oplus}$  depends on the unknown variables $c_4$ and $c_5$ and on whether the surface of the Earth lies in the region $(R_2/R_{\star})^3R_2<R_{\oplus}<R_{\star}$ or the region $0<R_{\oplus}<(R_2/R_{\star})^3R_2$, as this changes the form of $\pi^{\prime}_{\oplus}$.  In \cite{Nicolis:2008in} it was shown that the stability of a spherically symmetric field configuration, and subluminality of  fluctuations around this background require the conditions $c_4\geq 0$ and $c_5<0$.  Therefore even the sign of $Z_{\oplus}$ is not known from previous measurement meaning that the Galileon force between the plates of the experiment could be arbitrarily  attractive or repulsive depending on the relative magnitudes of $c_4$ and $c_5$. Laboratory experiments with parallel plates will constrain the Galileon parameter space through the combination of coefficients $Z_{\oplus}$.


\subsection{The Effect of the Cavity}

The laboratory experiments  we consider  are performed  inside a vacuum cavity in order to screen effects from the background environment.  Therefore it is necessary to check whether performing an experiment inside a vacuum cavity could affect the background Galileon field profile due to the Earth.

Firstly,  we find the solution for the cavity in isolation by neglecting the  background field profile due to the Earth.  As an approximation to the true experimental environment we consider a spherical vacuum cavity of radius $R$, surrounded by walls of thickness $\Delta$ and density $\rho$.  The equation of motion is
\begin{equation}
\frac{1}{r^2}\frac{\partial}{\partial r}\left[c_2\left(\frac{\pi^{\prime}}{r}\right)+2c_3\left(\frac{\pi^{\prime}}{r}\right)^2+2c_4\left(\frac{\pi^{\prime}}{r}\right)^3\right]=\rho\Theta(r-R)\Theta(R+\Delta-r)\;.
\end{equation}
For $r<R$ this has solution
\begin{equation}
\left(\frac{\pi^{\prime}}{r}\right)\left[c_2+2c_3\left(\frac{\pi^{\prime}}{r}\right)+2c_4\left(\frac{\pi^{\prime}}{r}\right)^3\right]=\frac{A}{r^3}\;,
\end{equation}
where $A$ is an arbitrary constant of integration.  To avoid divergences at the origin we choose $A=0$.
Therefore inside the vacuum chamber $\pi^{\prime}=0$.

For $R<r<R+\Delta$ the solution has the form
\begin{equation}
c_2\left(\frac{\pi^{\prime}}{r}\right)+2c_3\left(\frac{\pi^{\prime}}{r}\right)^2+2c_4\left(\frac{\pi^{\prime}}{r}\right)^3=\frac{\rho}{3}\left(1-\frac{R^3}{r^3}\right)+B\;,
\end{equation}
where $B$ is another arbitrary constant of integration.  Continuity of $\pi^{\prime}$ at $r=R$ imposes $B=0$. For distances $r\gg R$ the force $\pi^{\prime}$ tends towards the form
\begin{equation}
\pi^{\prime}\approx \alpha r\;,
\end{equation}
where $\alpha$ satisfies $c_2\alpha+2c_3\alpha^2+2c_4\alpha^3=\rho/3$. If the density of the material forming the walls of the cavity is of the order$ \mbox{ gcm}^{-3}$,  comparable to the density of the Earth, then the cavity produces a  Galileon force that is much smaller than that due to the Earth as long as the walls of the cavity have a thickness that is much smaller than  the radius of the Earth.  Clearly this is always the case in viable experimental scenarios.


As discussed in the previous section the variation in the background field due to the Earth over the metre lengths scales of the cavity is  extremely small.
Therefore, for an experimental environment of this sort, we can safely superimpose the cavity solution on to  the background field due to the Earth, where $\pi_{\oplus}^{\prime}$ is treated as constant, and its effect is to rescale  the coefficients.  This means that we can trust the form of the solution we have found for the cavity, even when it is placed in the vicinity of the Earth.

We can now study the Galileon force  between a pair of plates in an experiment.  We superimpose $\pi^{\prime}(\vec{x})= \pi^{\prime}_{\oplus}(R_{\oplus}) +\pi^{\prime}_{\rm cavity}+\phi^{\prime}_{\rm plates}(\vec{x})$, where $\pi^{\prime}_{\rm cavity}=0$, and $\phi_{\rm plates}^{\prime}$ is a  small perturbation to the Earth's contribution. Indeed the background force due to the Earth is a constant and acts to rescale  $c_2$, through the parameter $Z_{\oplus}$ as described in Section \ref{sec:Earth}, implying that at the cavity and plate level the theory behaves like a linear theory with a new coupling $Z_{\oplus}$.    From this point on we can study the Galileon field around the parallel plates of a laboratory experiment, free from considerations of the cavity or the  background configuration due to the Earth as long as we make the replacement
\begin{equation}
c_2\rightarrow Z_{\oplus}\;.
\end{equation}

\section{Experimental Constraints}
\label{sec:exp}
The experiments  we  consider consist of  two plates, aligned perpendicular to the $z$-direction, whose extent in the $x$,$y$-directions is much larger that their separation so we can approximate them as infinite.  The plates have density $\rho$.  The lower edge of one plate and the upper edge of the other is positioned at $z=d$ and $z=-d$ respectively, and the plates have width $\Delta$.  Therefore $\phi^{\prime}$, the strength of the Galileon force due to the configuration, is given by
\begin{equation}
\phi^{\prime}=\frac{\rho}{Z_{\oplus}}\left\{\begin{array}{lc}
z-d & d<z<d+\Delta\\
0 & -d<z<d\\
z+d & -(d+\Delta)<z<-d
\end{array}\right.\;,
\end{equation}
where we have imposed continuity of $\pi^{\prime}$ at the boundary of the plates, and $\pi^{\prime}(z)=-\pi^{\prime}(-z)$.
The approximation that the plates are infinite is valid whenever their extent in the $x$,$y$-directions is much larger than the distance $2d$ between the plates.

Taking this as the basic set up, we proceed to apply the results of laboratory experiments searching for deviations from
Newtonian gravity, and discuss the relevance of measurements of the Casimir force for Galileon searches.

\subsection{E\"{o}t-Wash}
The E\"{o}t-Wash experiment \cite{Kapner:2006si} looks for deviations from the Newtonian gravitational potential.  Two plates are suspended one above the other.  They have holes cut in them so that the area of overlap between the plates  changes as the plates are rotated. This experiment is an almost null experiment as the plates are designed in such a way that the torque due to Newton's force between the plates almost vanishes. The residual torque is then fitted with Newton's law  over distances between 55 microns and 9.53 mm. The discrepancy between Newton's law and the experimental results for small values of the plate distance sets a bound on the presence of non-Newtonian forces in this range of distances.
The  torque from any beyond the Standard Model physics is constrained to be $T<0.87\times 10^{-17}\mbox{ Nm}$. This will give us a bound on the coupling $\beta$.

The torque induced by the presence of the Galileon can be deduced from the energy of the plate-plate configuration. 
The energy in the top plate due to the perturbation of the Galileon field caused by the plates is
\begin{eqnarray}
E&=&A\int_d^{\Delta+d}dz\; Z_{\oplus}\phi\frac{d^2\phi}{dz^2}\;,\\
&=&A\int_d^{\Delta+d}dz\;\frac{\rho^2}{2Z_{\oplus}}(z-d)^2\;,\\
&=& \frac{A\rho^2\Delta^3}{6Z_{\oplus}}\;,
\end{eqnarray}
where $A$ is the area of overlap between the plates.  The torque  is the derivative of the energy of the top plate with respect to the angle of rotation.  So that the torque induced by the Galileon is
\begin{equation}
T=\frac{\rho^2\Delta^3}{6Z_{\oplus}}a_T\;,
\end{equation}
where $a_T=dA/d\theta$ is a constant which depends on the experimental setup
$a_T=3\times 10^{-3} \mbox{ m}^2$. The width of the plates is $\Delta = 1\mbox{ mm}$ and the plates are made of molybdenum with a density  $\rho = 10.28 \mbox{ gcm}^{-3}$.
The constraint on the Galileon force thus becomes
\begin{eqnarray}
Z_{\oplus}&>&6.05 \times 10^{40}\mbox{GeV}^2\;,\\
&>&(20m_P)^2\;,
\end{eqnarray}
which translates into a constraint on the coupling 
\begin{equation}
\beta < 0.05\;.
\end{equation}
Hence we find that in the context of the E\"{o}t-Wash experiment, the Galileon force between the plates must be much weaker than the gravitational one. This sets a bound on a previously unconstrained combination of the Galileon parameters.



\subsection{Casimir Experiments}
The Casimir force arises between two uncharged metallic plates in vacuum as a consequence of the quantisation of the electromagnetic field.  The ideal Casimir experiment directly measures the pressure on two parallel plates held extremely close together.  This would also be an ideal environment to study the Galileon force.
The average pressure on a plate due to the Galileon force is given by the integral of $\phi^{\prime}$ over the width of the plate.
\begin{eqnarray}
p=\frac{F}{A}&=&\int^{d+\Delta}_{d}dz \; \phi^{\prime}\rho\;,\\
&=& \frac{\rho^2}{Z_{\oplus}}\int^{d+\Delta}_{d}dz \; (z-d)\;,\\
&=&\frac{\rho^2\Delta^2}{2Z_{\oplus}}\;.
\label{eq:cas}
\end{eqnarray}

The practical difficulties of arranging such an experimental environment however, mean that experimental tests of the Casimir force are often indirect and no longer suitable as a test of the Galileon.  The best current measurement of the Casimir force by Decca et al. \cite{Decca:2007jq,Decca:2007yb,Klimchitskaya:2009cw} studies the gradient of the Casimir force between a plate and a sphere as a function of separation, whilst this is convenient for studying the Casimir force it is not suitable as a test of  the Galileon.  The sphere-plate experiment breaks the spatial symmetries which were vital in our calculation to ensure that the non-linear Galileon terms did not enter the equation of motion describing the scalar force.  If the non-linear terms are  present they act to additionally screen the force, as occurs inside the Vainshtein radius of a spherically symmetric system as discussed in Section \ref{sec:vainshtein}.  Additional screening of the sources means that the Galileon force is suppressed, and constraints from experiments are correspondingly weaker.     It is clear from Equation (\ref{eq:eom}) that   the screening effects of the non-linear terms can only be avoided in an experimental environment which varies in just one spatial dimension \footnote{Proposals to search for the Casimir effect using corrugated plates \cite{Bao:2010zz,PhysRevB.80.121402,PhysRevB.81.115417} are also unsuitable as tests of the Galileon, for the same reason that the additional spatial variation lead to screening effects, so the sensitivity of the experiment to the Galileon field is significantly reduced.}. Therefore we do not further analyse constraints from sphere-plate Casimir experiments here, as they are not  competitive with those from the E\"{o}t-Wash experiment.

The Casimir force has also been studied indirectly by measuring the gradient of the Casimir force as a function of the separation between two parallel plates \cite{Bressi:2002fr,0264-9381-18-18-313}.  Unfortunately this is also unsuitable as a test of the Galileon; as can be seen from Equation (\ref{eq:cas}) when the vacuum energy density is neglected the Galileon force is not a function of the separation between the plates, and therefore would not be seen in such an experiment.  Including the density of the vacuum reintroduces a dependence on the separation between the plates, but this experiment cannot be competitive with the E\"{o}t-Wash measurements in bounding the Galileon.

In \cite{Brax:2010xx} it was proposed that a modified parallel plate Casimir experiment could be used to search for the Chameleon by exploiting the change in the Chameleon force as the density of the inter-plate medium changes.  As the Galileon force also depends on the local energy density this experiment could also provide useful constraints on the Galileon model, an analysis that we leave for future study.

\section{Conclusions}
\label{sec:Conc}
Canonical scalar fields coupled to matter are  very tightly constrained by experimental searches for fifth forces.  One reason the Galileon model is interesting  is that it allows for such a coupling between a scalar field and matter, but the effects of this scalar are screened from searches for fifth forces by a dynamical mechanism.  The effects of the scalar field are suppressed within the Vainshtein radius, but they do not completely vanish, and so bounds can be placed on the parameters of the model using current laboratory experiments.  In addition we find that laboratory measurements constrain a combination of the four Galileon coefficients, called $Z_{\oplus}$ in this article, which is orthogonal to the constraints previously applied to the Galileon model.

\section*{Acknowledgments}
We would like to thank Justin Khoury, Claudia de Rham, Raquel Ribeiro, David Seery and Andrew Tolley for helpful discussion in the preparation of this work.  CB is supported by the  SNF. ACD is supported in part by STFC. CB and ACD wish to thank the theory group at CEA Saclay for hospitality whilst this work was initiated.

\section*{References}
	\bibliographystyle{JHEP}
\bibliography{galileon_resubmit}

\end{document}